\documentstyle[12pt]{article}
\begin{document}

\author{Victor Novozhilov
 and Yuri Novozhilov\footnote{E-mail  : yunovo@pobox.spbu.ru} \\
V.Fock Institute of Physics, \\
St.Petersburg State University, 198504, St.Petersburg, Russia}
\title{Color chiral solitons }

\date{}
\maketitle

\begin{abstract}
We discuss specific features of color chiral solitons (asymptotics,
possibility of confainment and quantization) at an example of isolated
SU(2) color skyrmions, i.e. skyrmions in a background field
which is the vacuum field forming the gluon condensate.

\end{abstract}

\vskip 6pt
{\bf Introduction}
\vskip 3pt

In low energy QCD the idea of flavor Skyrmions \cite{
Skyrme,Bala,Witten,Manton} was quite fruitful. Color chiral solitons and
skyrmions were studied less, although to find stable color configurations
could be an important step towards understanding of diquarks and exotic
hadrons. A possibility of color chiral solitons with baryon number $N_F/N_C$
was mentioned in the first paper on colour bosonization \cite{AAA+YuVN} .
However, the effective action in bosonization \cite{AAA+YuVN} was implicitly
gauge dependent, the choice of background colour fields was not discussed,
and soliton stability was not investigated. It was found \cite{YaF} that
direct application of an effective bosonized lagrangian \cite{AAA+YuVN} does
not lead to stable configurations. The idea of color skyrmions from
different viewpoints was explored \cite{Kaplan,Gomel,Karliner1, Karliner2}
in attempt to construct a constituent quark for $N_F=1$, but further
development in this direction was suspended after the conclusion \cite{
Karliner3} that stable colour solitons do not exist. Recently it was shown
that color solitons become stable due to background vacuum field which
should always be present with isolated soliton \cite{Novonovo, TMF}

In this talk we describe some specific feature of color solitons compared
with flavor solitons.

In QCD the flavor soliton action arises as a result of bosonization. The
chiral color bosonization in QCD follows , in general, the lines of flavor
bosonization \cite{AAA,Karchev,LMP,Ball} . The gluon field is a dynamical
gauge field, while the flavor field in the Dirac lagrangian is an external
one. In order to get a chiral color action, the background field should be
also chirally rotated giving an additional contribution to the standard
chiral action. In flavor bosonization no such terms are present \cite
{Novonovo}. This contribution is not considered here.

For a color soliton, the background field describes soliton environment and
produces corresponding interaction terms in the effective chiral lagrangian.
In this paper we consider a separate (free) soliton, and, therefore, take
color vacuum field as a background field. Such color vacuum fields form the
gluon condensate. Experimentally, the condensate is positive, so that the
vacuum field is chromomagnetic. Its value is a parameter of the theory.
Asymptotic behaviour of soliton configuration at large distances from the
centre is determined by the condensate: it is exponentially decreasing for
positive condensate and periodic otherwise.

In section 2 we start with general expression for a bosonization motivated
effective chiral lagrangian and work out an effective chiral action in the
case of the gauge group SU(2) and one flavor $N_F=1$ for the hedghog
configuration, study asymptotic behavior and evaluate the mass. In section 3
we define a two-soliton potential at large distances and show that it may
lead to confinement. In section 4 we briefly discuss quantization of color $%
SU\left( 2\right) $ skyrmion.

\vskip 6pt
{\bf Effective lagrangian for SU(2) soliton and choice of background
field}
\vskip 3pt

We write an  effective bosonization lagrangian for the color chiral field $U$
in the form, where we have omitted terms inessential for our discussion

$$
L_{eff}\left( U\right) =N_Ftr_C\{\frac{f_0^2}4D_\mu UD^\mu U^{-1}
$$
$$
+\frac 1{192\pi ^2}\left[ \frac 12\left[ UD_\nu U^{-1},UD_\mu U^{-1}\right]
^2-(UD_\nu U^{-1}UD^\nu U^{-1})^2\right] \}
,\eqno(1)
$$
As in bosonization, we write the kinetic term with a constant $f_0^2$ which
is an analogue of the pion decay constant $f_\pi ^2$. The first term in the
second line is the Skyrme potential, next term is specific for the
bosonization potential . We need it to show that it does not change general
properties of classical color soliton.  $D_\mu U=\partial _\mu U+\left[
G_\mu ,U\right] $ .

Color configurations are always associated with background color field $%
G_\mu $ because of necessity to maintain color gauge invariance. In this
respect, color solitons is quite different from flavor solitons, where there
is no flavor gauge invariance, and the external flavor gauge field can be
eliminated from the chiral action. We consider the color gauge group SU(2)
with antihermitian generators $T_a=\frac{\tau _a}{2i}$ , where $\tau _a $
are the Pauli matrices.

The background color field should be chosen according to the problem under
consideration. Our first step is to study a single colour soliton, i.e. a
soliton in the vacuum of gluonic field. The gluonic vacuum $\Psi _0$ is
characterized by the condensate
$$
C_g=\left( \Psi _0,\frac{g^2}{4\pi ^2}O_{\mu \nu }^aO^{\mu \nu a}\Psi
_0\right) \cong \frac{g^2}{4\pi ^2}G_{\mu \nu }^aG^{\mu \nu a}\neq 0\eqno(2)
$$
that is by the non-zero vacuum expectation value of the Yang-Mills
lagrangian for the full quantum field $O_\mu $ represented by the background
vacuum field $G_\mu $ in our approximation. According to phenomenological
descpription $C_g\succ 0$ , so that $G_\mu $ is a chromomagnetic field in
the real case of SU(3) gauge group. The vacuum field strength $G_{kl}$ in
the temporal gauge $G_0=0$ is constant up to a time independent gauge
transformation. The effective Lagrangian $L_{eff}$ is invariant under gauge
transformations of background fields and the chiral field.

We shall consider the simplest case of a chromomagnetic vacuum background
field , when it is an Abelian-type field which is a product a coordinate
vector field $V_k$ and a SU(2) color vector $n^a$
$$
G_k^a=V_kn^a,V_k=-\frac 12V_{kl}x_l=-\frac 12\varepsilon _{klm}x_l\nu
_mB,G_k=gG_k^a\frac{\tau _a}{2i}\eqno(3)
$$
where $n^a$ is a constant unit vector in the colour space, $\nu _m$ is a
constant unit vector in coordinate space, $\nu _mB=\frac 12\varepsilon
_{mlk}V_{lk}$ is the vacuum chromomagnetism and $B$ is related to the
condensate $C_g=\frac{g^2}{2\pi ^2}B^2$ . In the vacuum all directions $n^a$
and $\nu _l $ are equivalent, so that it is necessary to average over them
at the end. Such a choice of vacuum field does not lead to stability
troubles in QCD; although imaginary terms were detected at one-loop level
\cite{Savvidy}, they disappear in all-loop treatment \cite{Elizalde}.

Let us write the chiral field in the usual way
$$
U = \exp i\left( \frac{x_a\tau _a}R\right) F\left( R\right) =\cos F+ i{\bf r}%
\sin F, \\r_ar_a = r^2=1, r_a\tau _a={\bf r},r_a=\frac{x_a}R\eqno(4)
$$
Under a gauge transformation $S(\overrightarrow{x})$ the chiral field $U$
transforms together with the vacuum unit colour vector ${\bf n=}n_a\tau _a $
as
$$
U^{\prime }=SUS^{+},{\bf n}^{\prime }=S{\bf n}S^{+}+S\partial _kS^{+}
$$
and it is natural to restrict $S(\overrightarrow{x})$ by a condition $\left[
S,U\right] =0.$

Main structures in the Effective Chiral Lagrangian $L_{eff}\left( U\right) $
take on the following form
$$
D_kU = \partial _kU+g\frac{V_k}{2i}\left[ {\bf n},i{\bf r}\right] \sin F
$$
$$
S_{kl} = \left[ UD_kU^{+},UD_lU^{+}\right] =
$$
$$
= 4g\left( \left[ \overrightarrow{{\bf n}}, \overrightarrow{{\bf r}}\right]
_lV_k-\left[ \overrightarrow{{\bf n,}}\overrightarrow{{\bf r}} \right]
_kV_l\right) \frac{\sin ^2F} R+\partial _lU\partial _kU^{+}-\partial
_kU\partial _lU^{+}\eqno(5)
$$
where $R^2=x_kx_k$ . The kinetic structure $K$ and related non -Skyrme term $%
N$ are given by
$$
K = tr\left( D_lU^{+}D_lU\right) =2[\left( \partial _RF\right) ^2+\frac{
2\sin ^2F}{R^2}]+g^2V^2\left[ \overrightarrow{n},\overrightarrow{r}\right]
^2\sin ^2F\eqno(6)
$$
$$
N = tr\left( D_lU^{+}D_lU\right) ^2 =
$$
$$
= 2\left( \left( \partial _RF\right) ^2+\frac{2\sin ^2F}{R^2}\right) ^2+
\frac{16}{15}g^4V^4\sin ^4F+\frac 83g^2(\left( \partial _RF\right) ^2+ 3%
\frac{\sin ^2F}{R^2})V^2\sin ^2F\eqno(7)
$$
where $A^2=V_kV_k$ and we have averaged over directions ${\bf n}$ in the
SU(2) colour space putting $\overline{n_k}=0,\overline{n_k^2}=\frac 13,
\overline{n_k^4}=\frac 15,\overline{n_k^2n_l^2}=\frac 1{15}$ .

Similarly, we average over directions $\overrightarrow{\nu }$ of field $V_k$
in space of coordinates $x_k$ and get
$$
\overrightarrow{V}=\frac 12\left[ \overrightarrow{\upsilon },\overrightarrow{%
r}\right] RB,\overline{V^2}=\frac 16B^2R^2,\overline{V^4}=\frac 14\left(
1-\left( \overrightarrow{r,}\overrightarrow{\nu }\right) ^2\right)
_{av}^2B^4R^4=\frac 2{15}B^4R^4\eqno(8)
$$

It follows that the gauge field dependent part of the Skyrmion structure is
given by
$$
trS_{kl}S_{kl}-tr(S_{kl}S_{kl})_{B=0}=\frac{32}9g^2B^2\sin ^4F\eqno(9)
$$
while a mixed part of chirally transformed Lagrangian of the background
vacuum field acquires $\sin ^2F$
$$
trG_{lk}UG_{lk}U^{+}=-\frac 43g^2B^2\sin ^2F,G_{lk}=\frac g{2i}B\varepsilon
_{lkt}\nu _tn,trG_{lk}G_{lk}=-g^2B^2\eqno(10)
$$
We are now able to write down the Effective Color Static Lagrangian
$$
L_{eff}(U,G_k)=-N_F\frac{f_0^2}{16\pi ^2}[2(\left( \partial _RF\right) ^2+2%
\frac{\sin ^2F}{R^2})+\frac 29g^2B^2R^2\sin ^2F]
$$
$$
-\frac{N_F}{96\pi ^2}[\left( (\partial _RF)^2+2\frac{\sin ^2F}{R^2}\right)
^2+\frac{16}{225}g^4B^4R^4\sin ^4F+
$$
$$
+\frac 29g^2B^2R^2\left( (\partial _RF)^2+3\frac{\sin ^2F}{R^2}\right) \sin
^2F]-\frac{N_F}{48\pi ^2}\frac 23g^2B^2\sin ^2F
$$
$$
-\frac{N_F}{12\pi ^2}\left[ \frac{2\sin ^2F}{R^2}\left( \partial _RF\right)
^2+\frac{\sin ^4F}{R^4}+\frac 29g^2B^2\sin ^4F\right] \eqno(11)
$$
where terms with $gB$ arise from vacuum background field $G_\mu $ .

The Euler-Lagrange equation for (11) for the soliton function $F(R)$ has the
form
$$
N_Ff_0^2\left[ \left( 1+\frac{g^2B^2}6R^4\right) \sin 2F-2R\partial
_RF-R^2\partial ^2F\right] +
$$
$$
+\frac{N_F}{12\pi ^2}\left[ \frac{\sin ^2F\sin 2F}{R^2}-\sin 2F(\partial
_RF)^2-2\sin ^2F\partial ^2F+\frac 23g^2B^2R^2\sin ^2F\sin 2F\right] +
$$
$$
+\frac{N_F}{24\pi ^2}\left[ \frac{2\sin ^2F\sin 2F}{R^2}-2R(\partial
_RF)^3-3R^2(\partial _RF)^2\partial _R^2F-2\partial _R^2F\sin ^2F-(\partial
_RF)^2\sin 2F\right] +
$$
$$
+\frac{N_F}{675}g^4B^4R^6\sin ^2F\sin 2F
$$
$$
+\frac{N_F}{216\pi ^2}g^2B^2R^4\left[ \frac{3\sin ^2F\sin 2F}{R^2}-4\partial
_RF\frac{\sin ^2F}R-\partial _R^2F\sin ^2F\right]
$$
$$
-\frac{N_F}{432\pi ^2}g^2B^2R^4(\partial _RF)^2\sin 2F+\frac 13\frac{N_F}{%
24\pi ^2}g^2B^2R^2\sin 2F=0\eqno(12)
$$

This expression for the effective action contains terms $R^4\sin ^4F$ and $%
R^2\sin ^2F$ defining the asymptotic behaviour of $\sin F$ necessary to
obtain finite static energy or mass
$$
M=-4\pi \int dRR^2L_{eff}\left( U,G_k\right) \eqno(13)
$$
The contribution to the mass functional $M$ from the bosonized action sums
from the kinetic term, $d=4$ terms and the contribution of the background
vacuum field. It is easely to see that this part is positive definite and
bounded from below and provides with the soliton configuration.

We introduce dimensionless variable $\rho=Ef_0R$ , then the asymptotic
behavior at large $R$ of the decreasing function $F\left( \rho \right) $ is
represented by the following equation
$$
\partial _\rho \left[ \rho ^2\partial _\rho F\right] - 2(1+C\rho ^4)F=0, %
\eqno(14)
$$
where the dimensionless parameter $C=\pi^2Cg/9(Ef_0)^4$ is related to the
gluon condensate $Cg=g^2B^2/2\pi^2$. The solution of the Eq.(36) is modified
Bessel functions of the second kind $K\left( \frac34,\sqrt{\frac{C}{2}}%
\rho^2\right)/\sqrt\rho $ and asymptotic behavior
$$
F\rightarrow \rho ^{-\frac32}\exp \left( - \sqrt{\frac{C}{2}}\rho ^ 2\right)
,\rho\rightarrow \infty \eqno(15)
$$
which guarantees that the mass $M$ is finite.

At small $\rho$, as it can be expected, the soliton function $F(\rho )$
behaves near the origin $\rho =0$ in the same manner as in the Skyrme model $%
F\left( \rho \right) \approx \pi -b\rho $.

Thus, the function $F$ of the color soliton is quite different from that of
the flavor skyrmion.

We consider a family of trial functions
$$
F(\rho )=\pi \sqrt{\frac{1-b\rho +a\rho ^2}{1+A\rho ^5}}\exp (-\frac A2\rho
^2)\eqno(16)
$$
where coefficients a and b are variational parameters and parameter $A=\sqrt{%
2\pi ^2Cg/9(Ef_0)^4}$. We also minimize the mass functional with respect to
the scale transformation $\rho \rightarrow E\rho $. The functions (16)
reflects the behaviour at the origin and large distances (15). We look for
soliton configuration with $N_F=1$. We use the value for the gluon
condensate $C_g=(350MeV)^4$. We find stable soliton solutions for the wide
range of the unknown phenomenological parameter $f_0=(10...60)MeV$. When $f_0
$ corresponds to the mass scale $\Lambda _C=100MeV$ of the colour
bosonization, we get $M=460MeV$ .

\vskip 6pt
{\bf Two-soliton potential and possibility of confinement}
\vskip 3pt

Let us consider briefly main point leading to confining potential between
colour solitons in the case of $N_C=2,N_F=1$ discussed in section 3. For
such solitons, the vacuum background field plays the role of a bag. We take
two solitons $U\left( x_1\right) $ and $U\left( x_2\right) $ described by
Eq. (4) and ask what is an intersoliton potential ${\it V}\left(
x_{12}\right) $ at large distances $x_{12}$ and large $R_1,R_2$ , that
results from lagrangian (1) for two-soliton state $U(x_1,x_2;{\bf n,\nu }%
)=U\left( x_1,{\bf n,\nu }\right) U\left( x_2,{\bf n,\nu }\right) $ after
averaging over common colour and coordinate unit vectors ${\bf n,\nu }$ of
vacuum background field (3)

$$
{\it V}\left( x_{12}\right) F_1F_2=\left\langle H\left[ U\left( x_1,x_2;{\bf %
n,\nu }\right) \right] \right\rangle -\left\langle H\left[ U\left( x_1,{\bf %
n,\nu }\right) \right] \right\rangle -\left\langle H\left[ U\left( x_2,{\bf %
n,\nu }\right) \right] \right\rangle
$$
where $\left\langle H\right\rangle $ denotes averaging over ${\bf n,\nu }$
and $H$ is the static energy density following from (1). Second and third
terms in ${\it V}$ are given by $-L_{eff}$ in (11). The QCD part of the
effective lagrangian does not contribute. A gauge invariant potential
is
\begin{eqnarray*}
{\it V}\left( x_{12}\right) F_1F_2 &=&\frac{f_0^2}4\left\langle tr\left[
DU_2DU_1^{+}+DU_1DU_2^{+}\right] \right\rangle \cong tr\{\left[ G_k\left(
x_1\right) ,U_1\right] \left[ G_k\left( x_2\right) ,U_2^{+}\right] \} \\
{\it V}\left( x_{12}\right) &=&\frac{f_0^2}{36}g^2B^2\left( \overleftarrow{x_1}
-\overleftarrow{x_2}\right) ^2 \\
&&
\end{eqnarray*}
Such potential describes confined solitons. However, this does not explain
whether an isolated colour soliton can exist and propagate.

\vskip 6pt
{\bf Quantization of a color soliton}
\vskip 3pt

Quasiclassical quantization of solitons starts with introduction of
collective coordinates related to symmetries to be quantized. Method of
collective (group ) coordinates was first developed by N.N.Bogolyubov for
quantization of polaron. We follow approach considered by Balachandran and
al. \cite{Bala1} for flavor skyrmions. Quantization of color $SU\left( 2\right) $
skyrmion within color $SU\left( 3\right) $ was exposed by Kaplan \cite{Kaplan}. Time
development of static solution is associated with time dependent unitary
matrix $A\left( t\right) $ reflecting the symmetry of total (non-static)
lagrangian. In the color case \cite{Kaplan}$ A $ also a function of
$x_k $ and describes time development of color field as well.

For the soliton $U$ in the background field of quasiabelian type
$G_k^a\left( x\right) =v_k\left( x\right) N^a$ in the temporal gauge, color
gauge symmetry is reduced to the global one. Therefore, matrix $A\left(
t\right) $ can not depend on coordinated $x_k$ , and we introduce time
dependent soliton
$$
U\left( x,t\right)  = A\left( t\right) W\left( x\right) A^{+}\left( t\right)
\eqno(17) $$
Here $W\left( x\right) $ denotes the static chiral field configuration which
enters L$_{stat}$ as U$\left( x\right) $ .

Suppose now that the vacuum background field becomes time dependent too

$$
\hat G_k\left( x,t\right) =v_k\left( x\right) A\left( t\right) NA^{+}\left(
t\right) ,\hat G_0=0 , \eqno(18)
$$
remaining at the same time in the temporal gauge for $\hat G_\mu $ , while
$G_\mu $ gets $G_0=A^{+}\dot A$. The time development of $\hat G_k\left(
x,t\right) $

$$
i\partial _0\hat G_k=v_kA\left[ \left( A^{+}\dot A\right) ,N\right] A^{+}
\eqno(19)$$
depends on orientation of $\left( A^{+}\dot A\right) $ in the color space
with respect to the vacuum unit vector $N$ . The lagrangian of the gluonic
field $L_G$ will get a $G_0$ -dependent chromoelectric contribution
containing solitonic variables $A^{+}\dot A$ to be quantized
$$
\delta L_G=\frac 1{2g^2}tr\left[ G_0,G_k\right] ^2=-\frac{v_k^2}{2g^2}%
tr\left[ \left( iA^{+}\dot A\right) ,N\right] ^2   \eqno(20)
$$
and leading to a change of the gluon condensate. Thus, the assumtion that
background vacuum field becomes time dependent leads to inconsistency.
Therefore we require that within a domain vacuum direction $N$ is not
influenced by the unitary transformation $A\left( t\right) $
$$
A\left( t\right) NA^{+}\left( t\right) = N,A=\exp iNa\left( t\right) , $$
$$
\left( iA^{+}\dot A\right)  = N\dot a\left( t\right)  \eqno(21)
$$
so that $\left( iA^{+}\dot A\right) $ is also oriented in direction $N$. We
note that the one-loop Gauss law for the background field
$$
\frac{\delta L_G}{\delta G_0}=0,  \left[ N,\left[ N,\left( iA^{+}\dot
A\right) \right] \right] =0    \eqno(22)
$$
is satisfied for time-independent $N$.

Let us express the effective lagrangian $L_{eff}\left( U,\hat G\right) $ in
terms of new variables
$$
L_{eff}\left( U,\hat G\right)  = L_{stat}\left( W,G\right) +tr\left\{ \left[
\left( iA^{+}\dot A\right) ,W\right] \left[ \left( iA^{+}\dot A\right)
,W^{+}\right] \right\} + $$
$$
tr\left[ W^{+}[\left( iA^{+}\dot A\right)
,W],W^{+}D_lW\right] ^2   \eqno(23)
$$
In absence of background field and after $flavor \longleftrightarrow  color$
exchange, the lagrangian $L_{eff}\left( U,0\right) $ coincides with skyrmion
lagrangian in the flavor case. The Wess-Zumino action vanishes for
$SU\left( 2\right) $ soliton. Apart from $L_{stat}$ , the background field is
present only through $\left[ G_l,W\right] $ in $D_lW$ in the last term.

Integrating over coordinates we get the lagrangian for collective solitonic
variables in a particular vacuum domain with background vacuum field in
direction $N$
$$
\hat L\left( A,\dot A;N\right)  = -\frac 12\alpha \left( W,N\right) \left(
A^{+}\dot A\right) _iN_i\left( A^{+}\dot A\right) _jN_j    \eqno(24)
$$
where $\alpha $ is a numerical coefficient - ''momentum of inertia'' -
represented by space integrals over functionals of $W$ , as a solution of
variational equation for $L_{stat}\left( W,N\right) $.

Applying the quantization procedure we get the hamiltonian for the
soliton in a particular vacuum domain
$$
H=\frac{({\it R}_iN_i)({\it R}_iN_i)}{2\alpha _{}},  \eqno(25)
$$
where ${\it R}_k$ are $SU\left( 2\right) $ generators with commutation
relations
$$
\left[ {\it R}_i,{\it R}_k\right]  = -2ie_{ikj}{\it R}_j,\left[ {\it R}
_i,A_{kj}\right] =-\left[ A\sigma _i\right] _{kj},\left[
A_{ik},A_{jl}\right] =0 .
$$
Averaging over domains, i.e. over directions of domain vacuum fields,
gives us the hamiltonian for states of isolated soliton, as they are seen
''from outside''
$$
\hat H=\frac{{\it R}_i{\it R}_i}{6\alpha }     \eqno(26)
$$
Thus, averaging over vacuum domains reinstates the complete Hilbert space.
We refer to papers \cite {ANW} for its description in terms of spin and isospin.

\vskip 6pt
{\bf Conclusions and discussion}
\vskip 3pt

We have derived chiral colour action in a background field satisfying
standard conditions of one-loop QCD in background gauge. We applied this
action to the case of soliton configuration defined as a configuration in a
vacuum background field related to the gluon condensate, and considered the
static case. Experimentally, the condensate is positive, so that the vacuum
field is chromomagnetic. The vacuum field gives rise to terms in the
effective chiral lagrangian with $R^2\sin ^2F$ and $R^4\sin ^4F$ where $R$
is a distance from the center of the standard hedgehog configuration of the
shape $F.$ Vacuum background field ensures exponential decrease in
asymptotic region for chromomagnetic condensate, which is essential for
stability of soliton and finiteness of mass. The (renorm-invariant)
condensate is considered as a phenomenological quantity. Variational
estimates with the trial function with proper asymptotics behaviour and the
gluon condensate $(350Mev)^4$ shows that for the one flavor case the mass
cannot be more then 460 MeV. Color solitons definitely exist in bosonization
action and as Skyrmions.

Two-soliton potential displays confinement behavior. An essential element
leading to such potential is averaging over directions of vacuum fields in
different vacuum domains.

Quantization of collective color coordinates can be performed in the same
way as in the flavor case. Inside a particular vacuum domain the hamiltonian
of collective coordinates has cylindrical color symmetry determined by the
color vacuum field. Averaging this hamiltonian over vacuum domains we
recover spherical symmetry and an interpretation of spin- color isospin
states analogous to the flavor case.

Other topics of recent developments in color chiral solitons are discussed
in Ref \cite {Novonovo, TMF, Extend}.


\begin{thebibliography}{99}
\bibitem{Skyrme}  T.H.R.Skyrme, Proc.R.Soc.London A260 (1961) 127

\bibitem{Bala}  A.P.Balachandran, V.P.Nair, S.G.Rajeev and A.Stern,
Phys.Rev. D27 (1983) 1152, 2772

\bibitem{Bala1}  A.P.Balachandran, F.Lizzi, V.G.J.Rodgers and A.Stern,
Nucl.Phys. {\bf B256}, 525 (1985)

\bibitem{Witten}  E.Witten, Nucl.Phys.B223(1983) 422, 433

\bibitem{Manton}  N.S.Manton, Commun.Math.Phys. 111 (1987) 469

\bibitem{AAA+YuVN}  A.A.Andrianov and Yu.V.Novozhilov, Phys.Lett. B 153
(1985) 422

\bibitem{YaF}  V.A.Andrianov and V.Yu. Novozhilov, Sov.J.Nucl.Phys. 43
(1986) 626

\bibitem{Kaplan}  D.B.Kaplan, Phys.Lett. B235 (1990) 163; Nucl.Phys. B 351
(1991) 357

\bibitem{Gomel}  G. Gomelski, M.Karliner and S.B. Selipsky, Phys. Lett. B323
(1994) 182

\bibitem{Karliner1}  J.Ellis, Y.Frishman, A Hanany and M.Karliner, Nucl.
Phys. B382 (1992) 189

\bibitem{Karliner2}  Y.Frishman. A.Hanany and M. Karliner, Nucl. Phys. B424
(1994) 3

\bibitem{Karliner3}  Y.Frishman, A.Hanany and M.Karliner, hep-ph/9507206

\bibitem{Novonovo}  V.Novozhilov and Yu.Novozhilov, Phys.Lett. B 522
(2001) 49 ; hep-ph/0110006


\bibitem{TMF} V.Yu.Novozhilov and Yu.V.Novozhilov, Theor.Math.Phys. 131
(2002) 498

\bibitem{AAA}  A.Andrianov; Phys.Lett. 157B (1985) 425

\bibitem{Karchev}  N.Karchev and A.Slavnov, Teor.Mat.Fys. 65 (1985) 192

\bibitem{LMP}  A.A.Andrianov, V.A.Andrianov, V.Yu.Novozhilov,
Yu.V.Novozhilov Lett.Math.Phys. 11 (1986) 217

\bibitem{Ball}  R.D.Ball, Int.J.Mod.Phys.A, 5 (1990) 4391

\bibitem{Chaos}  V.Yu.Novozhilov and Yu.V.Novozhilov, J.Chaos 12 (2001)
2789; quant-ph/9911108

\bibitem{Savvidy}  G.K.Savvidy, Phys.Lett. 71B (1977) 133

\bibitem{Elizalde}  E.Elizalde and J.Soto, Nucl.Phys. B283 (1987) 577

\bibitem{ANW} G.Adkins. C.Nappi and E.Witten, Nucl.Phys. B 228 (1983) 552

\bibitem{Extend} V.Novozhilov and Yu.Novozhilov, hep-ph/0205175

\end{thebibliography}
\end{document}